\documentclass[twocolumn,prb,color,superscriptaddress,psfig]{revtex4}
\usepackage{graphicx}
\usepackage{epsfig}
\begin{document}
\title{Field induced antiferromagnetism and $^{17}$O Knight shift anomaly in La$_2$CuO$_4$}
\author{A.S.\ Moskvin}
\affiliation{Ural State University, 620083 Ekaterinburg,  
Russia}
\date{\today}
\begin{abstract}

We address the effect of the field induced antiferromagnetism in  paramagnetic state of the cuprate weak ferromagnet La$_2$CuO$_4$. The planar oxygen $^{17}$O Knight shift is shown to be an effective tool to inspect the effects of Dzyaloshinsky-Moriya coupling in cuprates in an external magnetic field. Field induced antiferromagnetism and anisotropic antiferromagnetic contribution to $^{17}$K explain the anomalies observed in $^{17}$O NMR in La$_2$CuO$_4$. The experimental observation of antiferromagnetic contribution to the $^{17}$O Knight shift provides probably the only way to find out the problem of the sense of Dzyaloshinsky vector in cuprates. 
 \end{abstract}

\maketitle


Starting from pioneer papers by Dzyaloshinsky \cite{Dzyaloshinsky} and Moriya \cite{Moriya} the 
Dzyaloshinsky-Moriya (DM) antisymmetric exchange coupling was extensively investigated in 60-80ths in connection with weak ferromagnetism focusing on  hematite $\alpha$-Fe$_2$O$_3$ and orthoferrites RFeO$_3$.\cite{Moskvin} The stimulus to a renewed interest to the subject was given by the cuprate problem, in particular, by the weak ferromagnetism observed in La$_2$CuO$_4$\cite{Thio} and  many other interesting effects for the DM systems, in particular, the "field-induced gap" phenomena \cite{Affleck} and field-induced staggered spin polarization.\cite{CuPM} It is worth noting that the latter effect was addressed earlier on in weak ferromagnet FeF$_3$.\cite{Moskvin2}

Below, in the paper we address the effect of the field-induced antiferromagnetism in La$_2$CuO$_4$ and its manifestation in the $^{17}$O NMR. We show that namely this effect explains  the puzzlingly  large negative $^{17}$O Knight shift for planar oxygens with anisotropy resembled that of weak ferromagnetism. \cite{Walstedt} 

  At variance with typical 3D systems such as orthoferrites or fluorides, cuprates  are characterised by a low-dimensionality, large diversity of Cu-O-Cu bonds including corner- and edge-sharing, different ladder configurations, strong quantum effects for $s=1/2$ Cu$^{2+}$ centers, and a particularly strong Cu-O covalency resulting in a comparable magnitude of hole charge/spin densities on copper and oxygen sites. 
 \begin{figure}[b]
\includegraphics[width=8.5cm,angle=0]{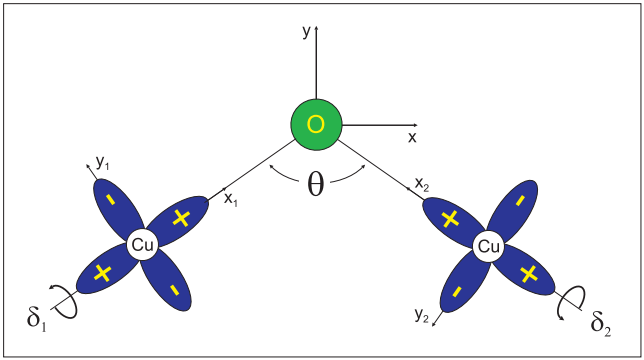}
\caption{Geometry of the three-center (Cu-O-Cu) two-hole system with ground Cu 3d$_{x^2-y^2}$ states.}
\label{fig1}
\end{figure}  
 
 We start with a typical for cuprates the three-center (Cu$_1$-O-Cu$_2$) two-hole system with tetragonal Cu on-site symmetry and ground Cu 3d$_{x^2-y^2}$ states (see Fig. 1) which conventional bilinear spin Hamiltonian is written in terms of copper spins as follows
\begin{equation}
\hat H_s(12)=J_{12}(\hat{\bf s}_1\cdot \hat{\bf s}_2)+{\bf D}_{12}\cdot [\hat{\bf s}_1\times \hat{\bf s}_2]+\hat{\bf s}_1{\bf \stackrel{\leftrightarrow}{K}}_{12} \,\hat{\bf s}_2 \, ,\label{1}
\end{equation} 
 where $J_{12}>0$ is an exchange integral, ${\bf D}_{12}$ is the Dzyaloshinsky vector, ${\bf \stackrel{\leftrightarrow}{K}}_{12}$ is a symmetric second-rank tensor of the anisotropy constants. 
 In contrast with $J_{12}, {\bf \stackrel{\leftrightarrow}{K}}_{12}$, the Dzyaloshinsky vector ${\bf D}_{12}$
 is antisymmetric with regard to the site permutation: ${\bf D}_{12}=-{\bf D}_{21}$. Usually this vector is assumed to be located on the bond connecting spins 1 and 2, though, strictly speaking, this should be written as a sum of three vectors located on Cu$_1$, Cu$_2$, and oxygen sites, respectively:${\bf D}_{12}={\bf D}^{(1)}_{12}+{\bf D}^{(O)}_{12}+{\bf D}^{(2)}_{12}$. Hereafter we will denote $J_{12}=J,{\bf \stackrel{\leftrightarrow}{K}}_{12}={\bf \stackrel{\leftrightarrow}{K}},{\bf D}_{12}={\bf D}$, respectively.  
 It should be noted that making use of effective spin Hamiltonian (\ref{1}) implies a removal of orbital degree of freedom that calls for a caution with DM coupling as it changes both a spin multiplicity,  and an orbital state.

   For a composite two $s=1/2$ spins system one should consider three types of the vector order parameters:
\begin{equation}
\hat{\bf S}=\hat{\bf s}_1+\hat{\bf s}_2;\,\hat{\bf V}=\hat{\bf s}_1-\hat{\bf s}_2;\,\hat{\bf T}=2[\hat{\bf s}_1\times \hat{\bf s}_2]
\end{equation}
with a kinematic constraint:
 \begin{equation}
 \hat{\bf S}^2+\hat{\bf V}^2=3\hat{\bf I};\,(\hat{\bf S}\cdot \hat{\bf V})=0;\,
(\hat{\bf T}\cdot \hat{\bf V})=6i; \, [\hat{\bf T}\times \hat{\bf V}]=\hat{\bf S}.
\label{SVT}
\end{equation}
In a sense the $\hat{\bf V}$ operator describes  the effect of local antiferromagnetic order, while $\hat{\bf T}$ operator may be associated with a vector chirality.\cite{Kolezhuk} In recent years, phases with broken vector chirality in frustrated quantum spin chains have attracted considerable
interest. Such phases are characterized by nonzero long-range correlations of the vector order parameter $\langle \hat{\bf T}\rangle $. Interestingly that  a chirally ordered phase can manifest itself as a "nonmagnetic" one, with 
$\langle \hat{\bf S}\rangle =\langle \hat{\bf V}\rangle =0$.

Both $\hat{\bf T}$ and $\hat{\bf V}$ operators  change the spin multiplicity with matrix elements  
$$
\langle 00|\hat T_m|1n\rangle =-\langle 1n|\hat T_m|00\rangle =i\delta_{mn};
$$
 \begin{equation}
\langle 00|\hat V_m|1n\rangle =\langle 1n|\hat V_m|00\rangle =\delta_{mn},
\end{equation}
 where we made use of Cartesian basis for $S=1$. The eigenstates of the operators  $\hat{V}_n$, and $\hat{T}_n$ with nonzero eigenvalues $\pm 1$ form N\'{e}el doublets $\frac{1}{\sqrt{2}}(|00\rangle \pm|1n\rangle$ and DM doublets $\frac{1}{\sqrt{2}}(|00\rangle \pm i|1n\rangle$, respectively. The N\'{e}el doublets correspond to classical collinear antiferromagnetic spin configurations, while the DM doublets correspond to quantum spin configurations which sometimes are associated with a rectangular 90$^0$ spin ordering in the plane orthogonal to the Dzyaloshinsky vector.

 Before going to microscopic analysis we should note that the interaction of our three-center system with external spins and/or fields $\hat H_{ext}$ is usually addressed by introducing only two types of effective external fields: the conventional Zeeman-like field and unconventional N\'{e}el-like staggered field, so that  $\hat H_{ext}$ reads as follows
 \begin{equation}
\hat	H_{ext}=-({\bf h}^S\cdot \hat {\bf S})-({\bf h}^V\cdot \hat{\bf V}).
\end{equation}
 It should be noted that an ideal N\'{e}el state is  attainable only in the limit of infinitely large staggered field, therefore for a finite staggered field ${\bf h}^V\parallel {\bf n}$ the ground state is a superposition of a spin singlet and a N\'{e}el state, 
$$
\Psi =\cos\alpha |00\rangle +\sin\alpha |1n\rangle,\, \tan2\alpha=\frac{2h^V}{J},
$$ 
 which composition reflects the role of quantum effects. For instance, in a Heisenberg spin 1/2 chain with $nn$ exchange the maximal value of staggered field $h^V=J/2$ hence the $\Psi$ function strongly differs from that of N\'{e}el state ($\langle \hat V_n\rangle =\sin2\alpha=\frac{1}{\sqrt{2}}$), and quantum mechanical average for a single spin $\langle s_z\rangle \leq\frac{1}{2}\sin\pi/4=\frac{1}{\sqrt{2}}\cdot\frac{1}{2}\approx 0.71\cdot\frac{1}{2}$ deviates strongly from classical value $\frac{1}{2}$. It should be noted that for the isolated antiferromagnetically coupled spin pair the zero-temperature uniform spin susceptibility turns into zero: $\chi ^S=0$, while for the staggered spin susceptibility we obtain $\chi ^V=2/J$.

 Application of an uniform external magnetic field ${\bf h}_S$ will produce a staggered spin polarization in the antiferromagnetically coupled Cu$_1$-Cu$_2$ pair
\begin{equation}
	\langle {\bf V}_{12}\rangle={\bf L}=-\frac{1}{J_{12}^2}[\sum_i{\bf D}_{12}^{(i)}\times {\bf h}^S]={\bf \stackrel{\leftrightarrow}{\chi}}^{VS}{\bf h}^S
\label{V}
\end{equation}
with antisymmetric $VS$-susceptibility tensor: $\chi _{\alpha\beta}^{VS}=-\chi _{\beta\alpha}^{VS}$. One sees that the sense of a staggered spin polarization, or antiferromagnetic vector, depends on that of  Dzyaloshinsky vector.\cite{Moskvin2}  The $VS$ coupling
results in many interesting effects for the DM systems, in particular, the "field-induced gap" phenomena in 1D s=1/2 antiferromagnetic Heisenberg system with alternating DM coupling.\cite{Affleck} Approximately, the phenomenon is described by a so called $staggered$  s=1/2 antiferromagnetic Heisenberg model with the Hamiltonian
\begin{equation}
\hat H=J\sum_i (\hat{\bf s}_i\cdot \hat{\bf s}_{i+1})-h_u\hat s_{iz}-(-1)^ih_s\hat s_{ix} \, ,\label{staggered}
\end{equation}  
which includes the effective uniform field $h_u$ and the induced staggered field $h_s\propto h_u$ perpendicular both to the applied uniform magnetic field and Dzyaloshinsky vector. 

Earlier on (see Ref.\onlinecite{Moskvin2}) we pointed to the ligand NMR as, probably, the only experimental technique to measure both staggered spin polarization, or antiferromagnetic vector in weak 3d-ferromagnets  and the value, direction, and the sense of Dzyaloshinsky vector. The latter possibility was realized with $^{19}$F NMR for  weak ferromagnet FeF$_3$.\cite{Moskvin2} 
Here we address the problem for generic cuprate weak ferromagnet La$_2$CuO$_4$.
 The DM coupling and magnetic anisotropy in La$_2$CuO$_4$ and related compounds
has attracted considerable attention in 90-ths (see, e.g., Refs.\onlinecite{Coffey,Koshibae,Shekhtman,debate}), and is still debated in the literature.\cite{Tsukada,Kataev} 
In the low-temperature tetragonal (LTT) and orthorhombic (LTO) phases of La$_2$CuO$_4$, the
oxygen octahedra surrounding each copper ion rotate by a
small tilting angle ($\delta _{LTT}\approx 3^0,\delta _{LTO}\approx 5^0$) relative to their location in the high-temperature tetragonal (HTT) phase.
The structural distortion  allows for the appearance of the antisymmetric Dzyaloshinsky-Moriya interaction.
In terms of our choice for structural parameters to describe the Cu$_1$-O-Cu$_2$ bond we have for LTT phase:
$ \theta =\pi; \delta _1=\delta _2=\frac{\pi}{2}\pm \delta _{LTT}$ for bonds oriented perpendicular to the tilting plane, and $ \theta =\pm(\pi -2\delta _{LTT}); \delta _1=\delta _2=\frac{\pi}{2}$ for bonds oriented parallel to the tilting plane. It means that all the local Dzyaloshinsky vectors turn into zero for the former bonds, and turn out to be perpendicular to the tilting plane for the latter bonds. For LTO phase:$ \theta =\pm(\pi -\sqrt{2}\delta _{LTO}); \delta _1=\delta _2=\frac{\pi}{2}\pm \delta _{LTO}$. The largest ($\propto \delta _{LTO}$) component of the local Dzyaloshinsky vectors (z-component in our notation) turns out to be oriented perpendicular to the Cu$_1$-O-Cu$_2$ bond plane. Other two components of the local Dzyaloshinsky vectors are fairly small: that of perpendicular to CuO$_2$ plane (y-component in our notation) is of the order of $\delta _{LTO}^2$, while that of oriented along Cu$_1$-Cu$_2$ bond
 axis (x-components in our notation) is of the order of $\delta _{LTO}^3$.

As an important by-product of the cuprate activity we arrived at a significant progress in different experimental methods and new opportunities to elucidate subtle details of electron and spin structure. In particular, it concerns the oxygen $^{17}$O NMR-NQR as an unique local probe to study the charge and spin densities on oxygen sites.
 In this connection we point to papers by R. Walstedt {\it et al}.\cite{Walstedt} as a first direct observation of  anomalous oxygen hyperfine interactions in  generic cuprate La$_2$CuO$_4$. With the approaching  transition to the ordered magnetic phase,  the authors observed anomalously large negative $^{17}$O Knight shift for planar oxygens which anisotropy resembled that of weak ferromagnetism in this cuprate. The giant shift was observed only when external field was parallel to the local Cu-O-Cu bond axis (PL1 lines) or perpendicular to CuO$_2$ plane. The effect was not observed  for PL2 lines which correspond to oxygens in the local Cu-O-Cu bonds which axis is perpendicular to in-plane external field.  The data were interpreted as an indication of oxygen spin polarization due to a local Dzyaloshinsky-Moriya antisymmetric exchange coupling. However, either interpretation of NMR-NQR data in such low-symmetry systems as La$_2$CuO$_4$ needs in a thorough analysis of transferred hyperfine interactions and a revisit of some textbook results  being typical for the model high-symmetry systems. First we draw attention to spin-dipole hyperfine interactions for O 2p-holes which are main participants of Cu$_1$-O-Cu$_2$ bonding. 
 Starting from a conventional formula for a spin-dipole contribution to local field
 $$
	{\bf \cal H}_n =-g_s\mu_B\sum_i\frac{3({\bf r}_i\cdot{\bf s}_i){\bf r}_i-r_i^2{\bf s}_i}{r_i^5}
$$
and making use of an expression for  appropriate  matrix element 
$$
	\langle p_i| \frac{3x_{\alpha}x_{\beta}-r^2\delta_{\alpha\beta}}{r^5}|p_j\rangle =-\frac{2}{5}\left\langle \frac{1}{r^3}\right\rangle _{2p}\langle p_i| 3\widetilde{l_{\alpha}l_{\beta}}-2\delta_{\alpha\beta}|p_j\rangle 
$$
\begin{equation}	
=	\frac{2}{5}\left\langle \frac{1}{r^3}\right\rangle _{2p}(\frac{3}{2}\delta_{\alpha i}\delta_{\beta j}+\frac{3}{2}\delta_{\alpha j}\delta_{\beta i}-\delta_{\alpha\beta}\delta_{ij})
\label{pipj}
\end{equation}  
we present a local field on the $^{17}$O nucleus in Cu$_1$-O-Cu$_2$ system as a sum of ferro- and antiferromagnetic contributions as follows\cite{Moskvin2} 
 \begin{equation}
{\bf \cal H}_n={\bf \stackrel{\leftrightarrow}{A}}^S\cdot \langle{\bf \hat S}\rangle+{\bf \stackrel{\leftrightarrow}{A}}^V\cdot \langle{\bf \hat V}\rangle
\end{equation}
where 
$$
{\bf \stackrel{\leftrightarrow}{A}}^S={\bf \stackrel{\leftrightarrow}{A}}^S(dp)+{\bf \stackrel{\leftrightarrow}{A}}^S(pd);\, {\bf \stackrel{\leftrightarrow}{A}}^V={\bf \stackrel{\leftrightarrow}{A}}^V(pd)-{\bf \stackrel{\leftrightarrow}{A}}^V(dp)\, ,
$$

$$
A^S_{ij}(dp)=A_p^{(0)}[3c_t(dp_i)c_t(dp_j)-|{\bf c}_t(dp)|^2\delta_{ij}]\, ,
$$
$$
A^S_{ij}(pd)=A_p^{(0)}[3c_t(p_id)c_t(p_jd)-|{\bf c}_t(pd)|^2\delta_{ij}]\, ,
$$ 
$$
A^V_{ij}(dp)=A_p^{(0)}[3\widetilde{c_s(dp_i)c_t(dp_j)}-({\bf c}_s(dp)\cdot{\bf c}_t(dp))\delta_{ij}]\, ,
$$
$$
A^V_{ij}(pd)=A_p^{(0)}[3\widetilde{c_s(p_id)c_t(p_jd)}-({\bf c}_s(pd)\cdot{\bf c}_t(pd))\delta_{ij}]\, ,
$$
where 
$$
c_{s,t}(dp_x)=-\frac{\sqrt{3}}{2}\frac{t_{dp\sigma}}{E_{s,t}(dp_x)}\sin\frac{\theta}{2}\,,
$$
\begin{equation} 
c_{s,t}(dp_y)=-\frac{\sqrt{3}}{2}\frac{t_{dp\sigma}}{E_{s,t}(dp_y)}\cos\frac{\theta}{2}
\end{equation} 
are probability amplitudes for different singlet ($c_s$) and triplet ($c_t$) 110 ($Cu_13d_{x^2-y^2}O2p_{x,y}$) and 011 ($O2p_{x,y}Cu_23d_{x^2-y^2}$) configurations in the ground state wave function, respectively; $c_{s,t}(dp_x)$=$-c_{s,t}(p_xd)$, $c_{s,t}(dp_y)=c_{s,t}(p_yd)$,  $t_{dp\sigma}$ is a hole dp-transfer integral, $A_{p}^{(0)}=\frac{2}{5}g_s\mu_B\left\langle \frac{1}{r^3}\right\rangle _{2p}$, the tilde points to a symmetrization. The energies $E_{s,t}(dp_{x,y})$  are those for singlet and triplet states of $dp_{x,y}$ configurations, respectively: $E_{s,t}(dp_{x,y})=\epsilon _{x,y}+K_{dpx,y}\pm I_{dpx,y}$, where $K_{dpx,y}$ and $I_{dpx,y}$ are Coulomb and exchange dp-integrals, respectively. Thus, along with a conventional textbook  ferromagnetic ($\propto \langle{\bf \hat S}\rangle$) transferred hyperfine contribution to local field which simply mirrors a sum total of two Cu-O bonds,  we arrive at an additional unconventional antiferromagnetic difference ($\propto \langle{\bf \hat V}\rangle$) contribution which symmetry and  magnitude strongly depend on the orientation of the oxygen crystal field axes and  Cu$_1$-O-Cu$_2$ bonding angle. In the case of Cu$_1$-O-Cu$_2$ geometry shown in Fig.1 we arrive at a diagonal ${\bf \stackrel{\leftrightarrow}{A}}^S$ tensor: 
$$
A^S_{xx}=2A_{p}(3\sin^2\frac{\theta}{2} -1);\,A^S_{yy}=2A_{p}(3\cos^2\frac{\theta}{2} -1);
$$
\begin{equation}
A^S_{zz}=-2A_{p},	
\end{equation}
and the only nonzero components of ${\bf \stackrel{\leftrightarrow}{A}}^V$ tensor: 
\begin{equation}
A^V_{xy}=A^V_{yx}=3A_{p}\sin\theta 	
\end{equation}
with
\begin{equation}
A_{p}=\frac{3}{4}\left(\frac{t_{dp\sigma}}{\epsilon_{p}}\right)^2A_{p}^{0}=f_{\sigma}A_{p}^{0},	
\end{equation}
where $f_{\sigma}$ is the parameter of a transferred spin density and we made use of a simple approximation $E_{s,t}(dp_{x,y})\approx \epsilon_{p}$. 
Generally speaking, we should take into account an additional contribution of magneto-dipole hyperfine interactions.

The two-term structure of oxygen local field implies a two-term  S-V structure of the $^{17}$O Knight shift
\begin{equation}
^{17}K={\bf \stackrel{\leftrightarrow}{A}}^S{\bf \stackrel{\leftrightarrow}{\chi}}^{SS} +
{\bf \stackrel{\leftrightarrow}{A}}^V{\bf \stackrel{\leftrightarrow}{\chi}}^{VS}	
\end{equation}
that points to Knight shift as an effective tool to inspect both uniform and staggered spin polarization. 
 The existence of antiferromagnetic term in oxygen hyperfine interactions yields a rather simple explanation of the $^{17}$O Knight shift anomalies in La$_2$CuO$_4$\cite{Walstedt} as a result of the external field induced  staggered spin polarization
$\langle{\bf \hat V}\rangle =\bf L = {\bf \stackrel{\leftrightarrow}{\chi}}^{VS} {\bf \cal H}_{ext}$. Indeed, "our" local $y$ axis for Cu$_1$-O-Cu$_2$ bond corresponds to  the crystal tetragonal $c$-axis oriented perpendicular to CuO$_2$ planes both in LTO and LTT phases of La$_2$CuO$_4$ while $x$-axis does to local Cu-O-Cu bond axis. It means that for the geometry of the experiment by Walstedt {\it et al.}\cite{Walstedt} (the crystal is oriented so that the external uniform field is either $\parallel$ or $\perp$ to the local Cu-O-Cu bond axis) the antiferromagnetic contribution to $^{17}$O Knight shift will be observed  only a) for oxygens in Cu$_1$-O-Cu$_2$ bonds oriented  along external field or b) for external field along tetragonal $c$-axis. Experimental data\cite{Walstedt} agree with staggered magnetization along the tetragonal $c$-axis in the former and along the rhombic $c$-axis (tetragonal $ab$-axis) in the latter.
Interestingly, the sizeable effect  has been observed in  La$_2$CuO$_4$ for temperatures $T\sim 500$ K that is essentially higher than $T_N \approx 300$ K.  Given $L=1$, $A_{p}^{(0)}\approx 100$ kG/spin,\cite{Walstedt} $|\sin\theta | \approx 0.1$, and $f_{\sigma}\approx 20\%$ we obtain $\approx 6$ kG as a maximal value of a low-temperature antiferromagnetic contribution to hyperfine field which is parallel to external magnetic field. This value agrees with a low-temperature extrapolation of high-temperature experimental data by Walstedt {\it et al.}\cite{Walstedt} Similar effect of anomalous $^{13}$C Knight shift has recently been observed in copper pyrimidine dinitrate  [CuPM(NO$_3$)$_2$(H$_2$O)$_2$]$_n$, a one-dimensional S=1/2 antiferromagnet with alternating local symmetry.\cite{CuPM}
However, the authors did take into account  only the inter-site  magneto-dipole contribution to ${\bf \stackrel{\leftrightarrow}{A}}^V$ tensor that  questions their quantitative conclusions regarding the "giant" spin canting.

The ferro-antiferromagnetic S-V structure  of local field on the nucleus of an intermediate oxygen ion in a  Cu$_1$-O-Cu$_2$ triad points to $^{17}$O NMR as, probably, the only experimental technique to measure both the value, direction, and the sense of Dzyaloshinsky vector. 
For instance, the negative sign of $^{17}$O Knight shift in La$_2$CuO$_4$\cite{Walstedt}  points to a negative sign of ${\bf \stackrel{\leftrightarrow}{\chi}}^{VS}$ for Cu$_1$-O-Cu$_2$ triad with $A_{xy}^V>0$, hence to a positive sense of $z$-component of  the summary Dzyaloshinsky vector in Cu$_1$-O-Cu$_2$ triad with geometry shown in Fig.1 given $\theta \leq \pi$, $\delta_1=\delta_2\approx \pi /2$.
It should be emphasized that the above effect is determined by the summary  Dzyaloshinsky vector in Cu$_1$-O-Cu$_2$ triad rather than by a local oxygen "weak-ferromagnetic" polarization as it was proposed by Walstedt {\it et.al.} \cite{Walstedt}

    In conclusion, we predict the effect of the field induced antiferromagnetism in a paramagnetic state of the cuprate weak ferromagnet La$_2$CuO$_4$. The planar oxygen $^{17}$O Knight shift is shown to be an effective tool to inspect the effects of Dzyaloshinsky-Moriya coupling in cuprates in an external magnetic field. Field induced antiferromagnetism and anisotropic antiferromagnetic contribution to $^{17}$K explain the anomalies observed in $^{17}$O NMR in La$_2$CuO$_4$.\cite{Walstedt} The experimental observation of antiferromagnetic contribution to the $^{17}$O Knight shift provides probably the only way to find out the problem of the sense of Dzyaloshinsky vector in cuprates.

 
 I thank R. Walstedt for stimulating and encouraging discussion. I would like to thank Leibniz-Institut f\"ur Festk\"orper- und Werkstoffforschung Dresden where part of this work was made for  hospitality. Partial support by  CRDF Grant No. REC-005, RFBR grants Nos. 04-02-96077, 06-02-17242, and 06-03-90893 is  acknowledged.

\end{document}